\documentclass[prl,twocolumn,preprintnumbers,amsmath,amssymb]{revtex4-1}
\usepackage{graphicx}

\begin{document}
\title{  Two indices  Sachdev-Ye-Kitaev model }
\author{ Jinwu   Ye }
\affiliation{
$^{1}$Department of Physics and Astronomy, Mississippi State University, MS, 39762, USA  \\
$^{2}$  Key Laboratory of Terahertz Optoelectronics, Ministry of Education and  Beijing Advanced innovation Center for Imaging Technology, Department of Physics, Capital Normal University, Beijing 100048, China  }
\date{\today }


\begin{abstract}
  We study the original Sachdev-Ye (SY) model in its Majorana fermion representation which
  can be called the two indices  Sachdev-Ye-Kitaev (SYK) model.
  Its advantage over the original SY model in the $ SU(M) $ complex fermion representation is that
  it need no local constraints, so a $1/M $ expansion can be more easily performed.
  Its advantage over the 4 indices SYK  model is that it has only two site indices
  $ J_{ij} $ instead of four indices $ J_{ijkl} $, so it may fit the bulk string theory better.
  By performing a $1/M $ expansion at $ N=\infty $, we show that a quantum spin liquid (QSL) state remains stable at a finite $ M $.
  The $ 1/M $ corrections are exactly marginal, so the system  remains conformably invariant at any finite $ M $.
  The 4-point out of time correlation ( OTOC ) shows quantum chaos neither at $ N=\infty $  at any finite $ M $, nor
  at $ M=\infty $ at any finite $ N $.
  By looking at the replica off-diagonal channel, we find there is a quantum spin glass (QSG) instability at
  an exponentially suppressed temperature in $ M $.
  We work out a criterion for the two large numbers $ N $ and $ M $ to satisfy so that
  the QSG instability may be avoided.
  We speculate  that at any finite $ N $, the quantum chaos appears at the order of $ 1/M^{0} $, which is the subleading order in
  the $ 1/M $ expansion.
  When  the $ 1/N $ quantum fluctuations at any finite $ M $ are considered, from a general reparametrization symmetry breaking point of view,
  we argue that   the effective action should still be described by the Schwarzian one, the  OTOC  shows maximal quantum chaos.
  This work may motivate future works to study the possible new gravity dual of the  2 indices SYK model.
\end{abstract}

\maketitle

{\bf 1. Introduction. }
    Sachdev-Ye(SY)  \cite{SY} studied the random $ SU(2) $ Heisenberg model  with infinite-range interactions:
\begin{equation}
  H_{H}= \frac{1}{ \sqrt{M}} \sum_{ij} J_{ij} \vec{S}_i \cdot \vec{S}_j
\label{SY}
\end{equation}
 where the random bond satisfies the Gaussian distribution $ P[ J_{ij} ] \sim e^{-NJ^2_{ij}/2 J^2 } $.

 In order to achieve some analytical results, SY generalized the $ SU(2) $ to $ SU(M) $ by
 introducing $ M $ complex fermions $ c_{i \alpha} $:
 $ S_{i;\mu,\nu} =c^{\dagger}_{i\mu} c_{i\nu} $ subject to the local constraint
 $ \sum_{\mu} c^{\dagger}_{i\mu} c_{i\mu} = q_0 M  $, then Eq.\ref{SY} becomes:
\begin{equation}
  H_{SY}=\frac{1}{ \sqrt{M}} \sum_{ij} J_{ij}  c^{\dagger}_{i\mu} c^{\dagger}_{j \nu} c_{i\nu}  c_{j\mu},~\sum_{\mu} c^{\dagger}_{i\mu} c_{i\mu} = q_0 M
\label{SYsum}
\end{equation}
    In the $ N \rightarrow \infty $ ( number of sites )  limit, followed by a $ M \rightarrow \infty $ limit in Eq.\ref{SYsum},
    SY found a gapless conformably invariant quantum spin liquid (QSL) ground state.
    At zero temperature $ T=0 $, the QSL has an extensive GPS entropy \cite{SY3,subir1,subir2}
    in the limit $ N \rightarrow \infty $ followed by  $ T \rightarrow 0 $  which is equal to the  Bekenstein-Hawking (BH) entropy
    in Einstein gravity  \cite{subir1,subir2,subir3}.

     In a series of talks in 2015, Kitaev \cite{Kittalk} simplified Sachdev-Ye model Eq.\ref{SYsum} to an infinite range four-indices Majorana fermion interacting model,
    each has $ N $ species:
\begin{equation}
H_{SYK}=  \sum^{N}_{i,j,k,l=1} J_{ijkl} \chi_{i} \chi_{j} \chi_k \chi_l
\label{SYK}
\end{equation}
    where  $  J_{ijkl}  $ also  satisfies the Gaussian distribution with
    $ \langle J_{ijkl} \rangle_J=0,  \langle J^2_{ijkl} \rangle_J= 3! J^2/N^3 $. By showing its possible maximal chaotic behaviour  matching the feat of the quantum black holes,
    Kitaev suggested that the $0+1 $ dimensional SYK  model may have a gravity
    dual in  asymptotic $ AdS_2 $ space.
    This speculation sparked great interests from both quantum gravity/string theory
    \cite{Pol,Mald,Gross,sff1,liu1,liu2,superSYK,randomsusy1,randomsusy2,u1zero,tensor1,tensor2,tensor3}
    and condensed matter/AMO community \cite{CSYKnum,Rcft,MBLSPT,tran1,tran2,highSYK1,highSYK3,highcSYK,longtime1,longtime2,rev}.
    Especially, Maldacena and Stanford did a systematic $ 1/N $ expansion \cite{Mald} on the SYK model.
    In the large $ N $ limit, it leads to the same gapless QSL ground state as that in the SY model.
    If dropping the irrelevant time derivative term $ \partial_\tau/J $, the saddle point equation ( and also the action ) has the time
    re-parametrization invariance $  \tau \rightarrow f( \tau) $, however,  the saddle point solution spontaneously breaks it to $ SL(2,R) $, leading to "zero mode " or Goldstone mode,
    while the irrelevant time derivative term explicitly breaks the re-parametrization symmetry and
    lifts the Goldstone mode to a pseudo-Goldstone mode whose quantum fluctuations
    can be described by  the Schwarzian action in terms of $ f( \tau) $ re-parametrization.
    From the Schwarzian, at the order $1/N $, they  evaluated the 4 point
    out of time ordered correlation (OTOC) function at early times   and extracted the Lyapunov exponent
    $ \lambda_L= 2 \pi/\beta $  at a small finite   temperature $ \beta J \gg 1 $.
    It is maximally chaotic and saturates the upper bound of some classes of quantum systems \cite{bound1,bound2,bound3}.
    This feat precisely matches that of quantum black holes in the Einstein gravity
    which are the fastest quantum information scramblers in the universe,
    therefore confirmed  the Kitaev's claim that the SYK model maybe
    dual to black holes in asymptotically $ AdS_2 $, which is, in fact,
    nearly conformably invariant/nearly $ AdS_2 $ with a scalar dilaton ( $ NCFT_1/NAdS_2 $ ).

  In this paper, we study the original Sachdev-Ye model in its Majorana fermion representation which can be called the two indices  Sachdev-Ye-Kitaev model.
  Its advantage over the original SY model in the $ SU(M) $ fermion representation Eq.\ref{SYsum} is that
  it need no local constraints. Its advantage over the 4 indices SYK model Eq.\ref{SYK} is that it has only two site indices $ J_{ij} $ instead of four indices $ J_{ijkl} $, so it may fit the bulk string theory better \cite{Pol}  .
  After the $ N \rightarrow \infty $ limit was taken, the $ 1/M $ expansion can be easily performed due to the absence of the constraint.
  It may also be easily generalized to short-range interaction in any space  dimension than the four indices SYK.
  By performing a $1/M $ expansion, we show that a quantum spin liquid (QSL) state remains stable at a finite $ M $.
  The $ 1/M $ corrections are exactly marginal,
  so they only change the values of the zero temperature entropy, the coefficient of the linear specific heat
  and the overall constants of all the  2 and 4 point  correlation functions.
  The system  remains conformably invariant at any finite $ M $.
  The 4-point out of time correlation ( OTOC ) shows quantum chaos neither at $ N=\infty $  at any values of $ M $,
  nor at $ M=\infty $ at any values of $ N $.
  Quantum chaos may only show up at a finite $ N $ and a finite $ M $ which can
  be explored by a $ 1/N $ expansion, followed by a  $ 1/M $ expansion.
  The two large numbers $ N $ and $ M $ play very different roles, $ N $ needs to be a large number
  to have a time window for quantum chaotic behaviours, therefore have a gravity dual in $ AdS_2 $,
  however, $ M $ also needs to be large enough to avoid the QSG phase,
  the $ 1/M $ expansion can only be used as a tool
  to evaluate the conformably invariant 2- or 4-point functions  or thermodynamic quantities at any finite $ M $.
  By looking at the replica off-diagonal channel, we find there is a quantum spin glass instability at
  an exponentially suppressed temperature $ T_{QSG} \sim J e^{ -\sqrt{ \pi M/2 } } $.
  We show that the QSG instability may be washed away when  $ M < N < e^{\sqrt{\pi/2 (M-1) }} $
  by the finite size effects at a finite $ N $.
  We argue that  when  the $ 1/N $ quantum fluctuations at any finite $ M $ are considered
  and if the QSG can be avoided,
  the effective action may still be described by the Schwarzian, the OTOC still show maximal quantum chaos.
  We expect the results achieved here also apply to the SY model which maybe called two indices complex fermion SYK,
  so it may also have a gravity dual in $ AdS_2 $ space, if QSG can be avoided.
  This work may inspire other works to study the possible new gravity dual of the 2 indices Majorana or Complex SYK model.



{\bf 2. Two indices SYK model. }
    Here we introduce a new class of SY model which can be named as two indices SYK model.
 Because $  SU(2)/Z_2= SO(3) $, there are two different ways to go to larger groups, one is generalize $ SU(2) $ to $ SU(M) $ as originally
 done by SY  in Eq.\ref{SYsum}.
 Here we take a different route, generalize $ O(3) $ to $ O(M) $. One can write a quantum spin
 in terms of $ M $ Majorana fermions $ \chi_{i \alpha}, \alpha=1,2....M $ at each site $ i $ :
\begin{equation}
  S^{\mu}_i = \frac{1}{2} \chi_{i\alpha} ( T^{\mu} )_{\alpha \beta} \chi_{i\beta}
\label{major1}
\end{equation}
 where  $ T^{\mu} $ is the $ M(M-1)/2 $ generators of the $ O(M) $ group \cite{om}.
 The $ M $ Majorana fermions satisfy the Clifford algebra $ \{\chi_{i \alpha}, \chi_{j \beta} \}= \delta_{ij} \delta_{\alpha \beta} $.
 For $ SO(3) $,  $ ( T^{\mu} )_{\alpha \beta}=-i \epsilon_{\mu \alpha \beta}  $.
 The total spin square $  \sum_{\mu}  ( S^{\mu}_i )^2= M(M-1)/8 $.
 Setting $ M=3 $ leads to the total spin $ s=1/2 $.
 Its main advantages over the original SY model Eq.\ref{SYsum} is that
 there are no constraints here.

 Using the Majorana fermion representation for a quantum spin has a long history:
 several authors including the author used it to solve multi-channel Kondo problems \cite{kondoye12345,stevekondo},
 Kitaev \cite{kit} and many others \cite{steveQSL} used a different
 version ( namely used 4 Majorana fermions by imposing an constraint ) to solve the
 quantum spin liquid (QSL) phase in a honeycomb lattice. Several authors used it to study QSL phases
 in anisotropic triangular lattices
 \cite{major1,major2}. This could be the first time to use it to solve a random quantum spin system.
 However, there are several tricky features  by using the Majorana fermions representation  of quantum spins
 which were noticed before \cite{major1,major2}:
 there is a $ Z_2 $ gauge degree of freedom $  \chi_{i\beta} \rightarrow -\chi_{i\beta} $ in Eq.\ref{major1},
 which played crucial roles in any description of QSL states.
 The Hilbert space of $ N $ spin $ 1/2 $ quantum spin is $ 2^{N} $,
 each spin $1/2 $ is represented by 3 Majorana fermions in the $ O(3) $ case,
 each Majorana fermion has quantum dimension $ \sqrt{2} $, so the Hilbert space of $3N $ of them
 is enlarged to $ 2^{N+[N/2]} $ where
 $ [N/2] $ takes the integer part of $ N/2 $. The extra $ 2^{[N/2]} $ dimension is due to the  $ Z_2 $ gauge degree of freedoms.
 We suspect that the physical consequences of this extra $ Z_2 $ degree of freedoms may increase the quantum fluctuations over the
 original quantum spins, therefore may favor quantum spin liquids over ordered states compared to the original quantum spin models.

  Substituting Eq.\ref{major1} to Eq.\ref{SY} leads to the two indices SYK model  written as SYK/2:
\begin{equation}
  H_{SYK/2} = \frac{1}{\sqrt{2M}} \sum_{ij} J_{ij} ( \chi_{i\alpha} \chi_{j\alpha} ) ( \chi_{i\beta} \chi_{j\beta} )
\label{SYmajor}
\end{equation}
 which, just like the SYK model Eq.\ref{SYK}, also contains 4 Majorana fermions, but with only  2 site indices $ ij $
 and additional $ O(M) $ index $ \alpha $. As argued below, may have several advantages over the original
 fermionic SY models in the $ SU(M) $ representation and the four indices SYK models.




 To solve the original $ SU(M) $ fermions SY model Eq.\ref{SYsum}, one need to take
 $ N \rightarrow \infty $ first, then followed by $ M \rightarrow \infty $.
 One must also introduce a Lagrangian multiplier to enforce the local constraint in Eq.\ref{SYsum} which becomes
 a global constraint at $ M=\infty $.
 Fixing at $ N= \infty $, the $ M \rightarrow \infty $ leads to the gapless QSL.
 As said in the introduction,  all these main difficulties of the original SY model were ingeniously circumvented by Kitaev by replacing the $ SU(M) $ fermions with $ N $ Majorana fermions, the two indices $ J_{ij} $ by 4 indices $ J_{ijkl} $,
 double large $ N $, large $ M $ limit by just one large $ N $ limit.
 This is a significant improvement over the original SY model both analytically and numerically.
 Here, we still keep the 2 indices $ J_{ij} $, replacing the $ SU(2) $ fermions by 3 Majorana fermions
 to keep the spin algebra at $ SU(2)/Z_2=SO(3) $, then extend the $ SO(3) $ to $ SO(M) $,
 still use the $ 1/N $ expansion, followed by a $ 1/M $ expansion.

 One of the biggest advantages of this $ O(M) $ group over the  $ SU(M) $ group is  the absence of
 a Lagrangian multiplier to enforce the local constraint in Eq.\ref{SYsum}.
 This make the following $ 1/M $ expansion much easier to perform than that of  $ SU(M) $.
 The advantage over the 4 indices SYK model \cite{Kittalk}  is that here we still stick to the two indices $ J_{ij} $.
 As argued in \cite{Pol},  two index coupling $ J_{ij} $ may fit better with a bulk string theory.

{\bf 3. The mean field solution at $ N = \infty $ followed by the  $ M =\infty $. }
 By using replica $ a, b =1,2,... n $ where $ n $ is the number of replicas,
 doing quenched average over the  Gaussian distribution $ P[J_{ij}] $ and introducing
 the Hubbard-Stratonovich (HS) field
 $ Q^{ab}_{\alpha \beta, \gamma \delta } ( \tau, \tau^{\prime} ) = Q^{ba}_{\gamma \delta, \alpha \beta} (\tau^{\prime},  \tau ) $,
 different sites are decoupled:
 \begin{eqnarray}
   \bar{ Z^n } = \int {\cal D} Q exp[ - N {\cal F}(Q) ] ~~~~~~~~~~~~   \nonumber   \\
 {\cal F}(Q)=  \frac{1}{J^2 M } \int d \tau  d \tau^{\prime} [ Q^{ab}_{\alpha \beta, \gamma \delta } ( \tau, \tau^{\prime} )]^2
 - \log Z_0
 \label{zn}
\end{eqnarray}
  where $ Z_0 $ is the single site partition function:
\begin{eqnarray}
    Z_0 & = & \int {\cal D} \chi exp[- \frac{1}{2} \int d \tau \chi^a_{\alpha} \partial_\tau \chi^a_{\alpha}
      +  \frac{1}{M} \int d \tau  d \tau^{\prime} Q^{ab}_{\alpha \beta, \gamma \delta } ( \tau, \tau^{\prime} )    \nonumber  \\
       & \cdotp & \chi^a_{\alpha} ( \tau) \chi^a_{\beta} ( \tau) \chi^b_{\gamma} ( \tau^{\prime} ) \chi^b_{\delta} ( \tau^{\prime}) ]
\label{z0}
\end{eqnarray}

    In the $ N \rightarrow \infty $ limit,  we get the following saddle-point equation for the $ Q $ field:
\begin{equation}
    Q^{ab}_{\alpha \beta, \gamma \delta } ( \tau, \tau^{\prime} )=
    \frac{J^2}{2} \langle \chi^a_{\alpha} ( \tau) \chi^a_{\beta} ( \tau) \chi^b_{\gamma} ( \tau^{\prime} ) \chi^b_{\delta} ( \tau^{\prime}) \rangle
\label{saddle1}
\end{equation}
   We assume that in both quantum spin liquid and quantum spin glass phase
   $  Q^{ab}_{\alpha \beta, \gamma \delta } ( \tau, \tau^{\prime} )
   = Q^{ab} ( \tau, \tau^{\prime} ) [ \delta_{\alpha  \delta } \delta_{\beta \gamma} - \delta_{\alpha \gamma  } \delta_{\beta \delta} ] $
   which is obviously anti-symmetric in $ (\alpha \beta) $ and $ (\gamma \delta ) $.

   In a  quantum spin-glass, $  Q^{aa}( \tau- \tau^{\prime} \rightarrow \infty )= q_{EA} \neq 0 $ which is the Edward-Anderson (EA) order parameter \cite{SY,rotor1,rotor2}. For its replica off-diagonal  $ a \neq b $ component $  Q^{a \neq b}( \tau- \tau^{\prime}  )= q \neq 0 $
   which is independent of  $ \tau- \tau^{\prime} $.  If the replica symmetry is not broken in the QSG phase, then $ q_{EA}= q $.


  Introducing a second HS field $ P_{ab}( \tau, \tau^{\prime} )= -P_{ba}( \tau^{\prime}, \tau ) $, one can transform
  $ Z_0 $ into the following form:
\begin{eqnarray}
   Z_0 & = & \int {\cal D} P exp[ - M {\cal F}_Q[ P ] ]      \nonumber   \\
   {\cal F}_Q[ P ] & = & 2 \int d \tau  d \tau^{\prime}  Q^{ab}( \tau, \tau^{\prime})
   P^2_{ab}( \tau, \tau^{\prime} ) -\log Z_{00}
\label{z00P}
\end{eqnarray}
  where $ Z_{00} $ is the single-site and single-component partition function:
\begin{eqnarray}
    Z_{00} & = &  \int {\cal D} \chi exp[- \frac{1}{2} \int d \tau \chi^a_{\alpha} \partial_\tau \chi^a_{\alpha}
      + 4 \int d \tau  d \tau^{\prime} Q^{ab} ( \tau, \tau^{\prime} )    \nonumber  \\
      & \times & P_{ab}( \tau, \tau^{\prime} )
      \chi^a_{\alpha} ( \tau) \chi^b_{\alpha} ( \tau^{\prime}) ]
\label{z00}
\end{eqnarray}

    In the $ M \rightarrow \infty $ limit, we reach the saddle-point equation for the two-point function:
\begin{equation}
 P_{0ab}( \tau, \tau^{\prime} ) = G_{0ab}( \tau- \tau^{\prime} )= \frac{1}{M}
 \langle \chi^a_{\alpha} ( \tau) \chi^b_{\alpha} ( \tau^{\prime}) \rangle
\label{saddle2}
\end{equation}

  In the $ M \rightarrow \infty $ limit,  Eq.\ref{saddle1} becomes:
\begin{equation}
Q^{ab}_0 ( \tau- \tau^{\prime} )= \frac{J^2}{2} G^2_{0ab}( \tau- \tau^{\prime} )
\label{saddle12}
\end{equation}

  From Eq.\ref{z00}, one can identify the system's self-energy:
\begin{equation}
 \Sigma_{0ab} ( \tau- \tau^{\prime} )= 4 J^2 G^3_{ab}( \tau- \tau^{\prime} )
\label{self}
\end{equation}
   and reach the following self-consistent equation:
\begin{equation}
  G_{0ab} ( i \omega_n )=( -i \omega_n - \Sigma_{0ab}( i \omega_n ) )^{-1}
\label{selfeq}
\end{equation}
 where the matrix inversion is taken in the replica space.


 Obviously, due to the fermions can not condense, so $ G_{0ab}( \tau- \tau^{\prime} ) =0 $ for $ a \neq b $.
 So the fermion Green function only has the replica diagonal saddle point solution, there is no QSG at $ M= \infty $.
 So in the following, we focus on the quantum spin-liquid phase.
 The possible instability to the QSG order will be discussed in the conclusion section.
 Then the self-consistent equations \ref{self},\ref{selfeq} for a single replica
 take the identical form as the SY model in the $ SU(M) $ representation \cite{SY,SY3} and the SYK model
 in the $ N=\infty $ limit  \cite{Kittalk,Pol,Mald,Gross}.
 So if dropping the irrelevant term $ \partial_{\tau} $ in Eq.\ref{selfeq},
 the saddle point equations \ref{self},\ref{selfeq} have parametrization invariance \cite{spin} under $ \tau \rightarrow f (\tau) $:
\begin{eqnarray}
   G( \tau_1, \tau_2 ) & \rightarrow &  [ f^{\prime}( \tau_1) f^{\prime}( \tau_2) ]^{\Delta} G( f(\tau_1), f(\tau_2) )
        \nonumber   \\
   \Sigma( \tau_1, \tau_2 ) & \rightarrow &  [ f^{\prime}( \tau_1) f^{\prime}( \tau_2) ]^{\Delta (q-1)}
   \Sigma( f(\tau_1), f(\tau_2) )
\label{ftau}
\end{eqnarray}
  where $  \Delta =1/q $ with $ q=4 $.

  The conformably invariant solution at a long time  was found to be ( after replacing $ J^2 $ in \cite{Kittalk,Pol,Mald,Gross}  by $ 4 J^2 $ ):
\begin{equation}
  G_{0}( \tau )= \frac{ \Lambda }{ |\tau|^{1/2} } sgn( \tau ),~~~\Lambda
  = (\frac{1}{ 16 \pi J^2 } )^{1/4}
\label{break}
\end{equation}
  which breaks the parametrization symmetry in Eq.\ref{ftau} down to the $ SL(2,R)$.

{\bf 4.  $ 1/M $ expansion at $ N=\infty $. }
  Fixing at $ N=\infty $, the saddle point Eq.\ref{saddle1} still holds.
  However, the saddle point Eq.\ref{saddle2} suffers quantum fluctuations. In performing the $ 1/M $ expansion
  at a fixed  $ Q^{ab} ( \tau, \tau^{\prime} ) $, it is convenient to use the self-energy
  $ \Sigma^{ab}= 8 Q^{ab} ( \tau, \tau^{\prime} ) P_{ab}( \tau, \tau^{\prime} ) $ to replace
  $ P_{ab}( \tau, \tau^{\prime} ) $, then Eq.\ref{z00P} becomes \cite{jacob}:
\begin{equation}
 {\cal F}_Q[ \Sigma ]= 2 \int d \tau  d \tau^{\prime}  \frac{ \Sigma^{2}_{ab}( \tau, \tau^{\prime})} {32
   Q_{ab}( \tau, \tau^{\prime} ) } -\log Pf ( \partial_{\tau} - \Sigma_{ab} )
\label{FQ}
\end{equation}
  where $  Q_{ab}( \tau, \tau^{\prime} ) $ should be  taken as a fixed external potential.
  Obviously taking the saddle point $ \frac{ \partial {\cal F}_Q[ \Sigma ] } { \partial \Sigma }=0 $
  recovers Eq.\ref{self}, \ref{selfeq}. At a finite $ M $, one can write:
\begin{equation}
 \Sigma_{ab} (  \tau, \tau^{\prime} )= \Sigma_0 (  \tau- \tau^{\prime} ) \delta_{ab} +
 \delta \Sigma_{ab}(  \tau, \tau^{\prime} )
\label{deltaSigma}
\end{equation}

\begin{figure}[tbp]
\includegraphics[width=8cm]{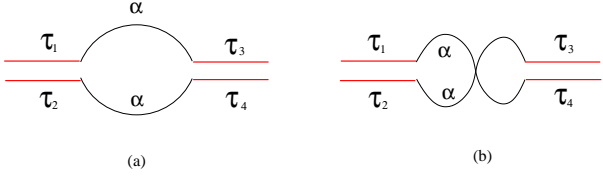}
\caption{(Color online) The quantum fluctuations of the self-energies (a) and (b) differs by a minus sign.
 The red bi-local double line stands for the $ \delta \Sigma $, the black solid line is the propagator of the Majorana fermions with the $ O(M) $
 spin index $ \alpha $ ( no sum over $ \alpha $ ). (a)+(b) leads to the second term in Eq.\ref{detlaS}.  }
\label{SYKpol}
\end{figure}

  In principle, when performing the $ 1/M $ expansion, one need to keep the saddle point Eq.\ref{saddle1}
  and solve it self-consistently order by order \cite{rotor1} in $ 1/M $.
  Fortunately,  to evaluate N-point correlation functions at the order of $ 1/M $,
  one can simply ignore the self-consistency Eq.\ref{saddle1} and set
  $ Q^{ab} ( \tau, \tau^{\prime} )= Q^{ab}_0 ( \tau- \tau^{\prime} ) $
  ( However, as to be shown later, this is not true in  evaluating $ 1/M $ corrections to the free energy ).

   In the following, we will ignore the replica off-diagonal $ a \neq b $ fluctuations, so only focus on
   the replica diagonal ones $ \delta \Sigma_{aa}(  \tau, \tau^{\prime} ) $ ( so we will drop the replica index $ a $ ).
   Substituting Eq.\ref{deltaSigma} into Eq.\ref{FQ}, one can see
   that the linear term vanishes, the quadratic term becomes:
\begin{eqnarray}
  F_{Q_0}[  \delta \Sigma ] & = & \int d\tau_1 d\tau_2  \frac{  ( \delta \Sigma( \tau_1, \tau_2 ) )^2 }{ 16 J^2 G^2_0( \tau_1 - \tau_2 ) }
    +  \frac{1}{4} \int d\tau_1 d\tau_2  d\tau_3 d\tau_4    \nonumber   \\
     & \times & \delta \Sigma( \tau_1, \tau_2 ) \delta \Sigma( \tau_3, \tau_4 )
     G_{0}( \tau_1 - \tau_3 ) G_{0}( \tau_2 - \tau_4 )
\label{detlaS}
\end{eqnarray}

   It is instructive to  see that the first term ( or the first term in Eq.\ref{FQ} ) coming from
   the combination of two HS fields $ Q $ and $ P $  is diagonal in  $ ( \tau_1, \tau_2 ) $ space,
   the Green function $ G_0( \tau_1 - \tau_2 ) $ appears in the denominator,  in the long time limit
   $ 1/G^2_0( \tau_1 - \tau_2 ) \sim  | \tau_1-\tau_2 | $ which diverges linearly \cite{naive}.
   However, the second term ( or the second term in Eq.\ref{FQ} ) coming from the integrations of
   the Majorana fermion bubbles ( Fig.1 )
   is off-diagonal in  $ ( \tau_1, \tau_2 ) $ and  $ ( \tau_3, \tau_4 ) $ space
   ( but they become  diagonal in the imaginary frequency space ),
   the Green functions $ G_0( \tau_1 - \tau_3 ) G_{0}( \tau_2 - \tau_4 )  $ appear in the numerator,  in the long time limit,
   the product $  |\tau_1 - \tau_3 |^{-1/2}  | \tau_4 - \tau_2 |^{-1/2} sgn ( \tau_{13} ) sgn ( \tau_{24} ) $
   decay to zero in the long time limit.
   However, due to the completely different dependencies of the two terms on the Green function,
   the first term dominates over the second, so  $ F_{Q_0}[  \delta \Sigma ] $ remains positive definite.
   It shows the stability of the QSL phase at least to the order of $ 1/M $.
   We expect it to be stable to all orders of $ 1/M $.
   It is also easy to see the $ 1/J $ in Eq.\ref{break} factors out, points to the conformably invariant
   form of $ F_{Q_0}[  \delta \Sigma ] $ in the long time limit.
   In fact, the first term is invariant under the following scale transformation:
   $ \tau_1 \rightarrow \lambda \tau_1, \tau_2 \rightarrow \lambda \tau_2 $,
   one knew $ 1/G^2_0( \tau_1 - \tau_2 ) \sim  | \tau_1-\tau_2 |  \rightarrow \lambda | \tau_1-\tau_2 | $,
   then if one assumes $ \delta \Sigma( \lambda\tau_1, \lambda\tau_2 ) \rightarrow \lambda^{-3/2} \delta\Sigma( \tau_1, \tau_2 ) $, then
   it indicates  $ \delta\Sigma( \tau_1, \tau_2 ) \sim  1/ | \tau_1-\tau_2 |^{3/2} $ which takes the same scaling form
   as the saddle point $ \Sigma_0( \tau_1, \tau_2 ) \sim  1/ | \tau_1-\tau_2 |^{3/2} $.
   Similarly, one can check the second term is invariant under the same scale transformation:
   $ \tau_i \rightarrow \lambda \tau_i, i=1,2,3,4 $, so Eq.\ref{detlaS} indicates $ \delta\Sigma( \tau_1, \tau_2 ) \sim  1/ | \tau_1-\tau_2 |^{3/2} $ which will be confirmed by a direct Feymann diagram calculation in Fig.2a and Eq.\ref{S1M}.
   In fact, one can check that the next order ( the sixth order ) term $ \int d\tau_1 d\tau_2  d\tau_3 d\tau_4  d\tau_5 d\tau_6
     \delta \Sigma( \tau_1, \tau_2 ) \delta \Sigma( \tau_3, \tau_4 ) \delta \Sigma( \tau_5, \tau_6 )
     G_{0}( \tau_2 - \tau_3 ) G_{0}( \tau_4 - \tau_5 ) G_{0}( \tau_6 - \tau_1 ) $ is also invariant under the  same scale transformation.



   In fact, one may make Eq.\ref{detlaS} physically more transparent by defining
   $\delta \Sigma( \tau_1, \tau_2 ) = 4 J G_0( \tau_1 - \tau_2 ) \delta \sigma( \tau_1, \tau_2 ) $, then Eq.\ref{detlaS} can be re-written as:
\begin{eqnarray}
  F_{Q_0}[  \delta \sigma ] & = &  \int d\tau_1 d\tau_2  ( \delta \sigma( \tau_1, \tau_2 ) )^2
    +   \int d\tau_1 d\tau_2  d\tau_3 d\tau_4    \nonumber   \\
     & \times & \delta \sigma( \tau_1, \tau_2 ) \delta \sigma( \tau_3, \tau_4 )
     K_{1/M} ( \tau_1, \tau_2; \tau_3, \tau_4 )
\label{detlas2}
\end{eqnarray}
 where $ K_{1/M} ( \tau_1, \tau_2; \tau_3, \tau_4 )  = 4 J^2 G_{0}( \tau_1 - \tau_2 ) G_{0}( \tau_1 - \tau_3 )
   G_{0}( \tau_2 - \tau_4 ) G_0( \tau_3 - \tau_4 ) $
   is identical to the kernel of the ladder diagram of the 4-point function in the SYK model \cite{Mald,kernel,absolute}.
   The eigenvalues and eigen-functions of the Kernel have been worked out in \cite{Mald} using the conformal invariance.
   By taking into account the replacement $ J^2 \rightarrow 4 J^2 $ in Eq.\ref{break}, one can see
   that kernel has a positive eigenvalue $ k_c(h)=\frac{\tanh \frac{\pi}{2} s }{ 2 s}  $ for the continuous conformal weight $ h=1/2+is $,
   negative eigenvalue $ k_c(h)=-\frac{1}{4n-1} $
   for the discrete conformal weight $ h=2n, n=1,2,3,\cdots  $.
   In both cases, Eq.\ref{detlas2} is positive definite.

  We also solve Eq.\ref{self},\ref{selfeq} numerically just like in \cite{SY} which recover the conformally invariant
  solution only in the long time limit,
  then plug them into Eq.\ref{detlaS} to show it remains positive definite when using the complete solutions.


\begin{figure}[tbp]
\includegraphics[width=8cm]{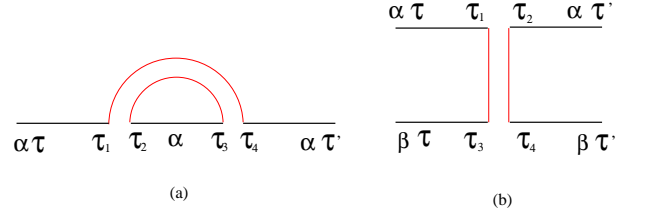}
\caption{(Color online) The $ 1/M $ correction to (a) the self energy and (b) the four point function $ Q^{ab} ( \tau- \tau^{\prime} ) $.
The red bi-local double line stands for the propagator of $ \delta \Sigma $, the black solid line is the propagator of the Majorana fermion with the $ O(M) $
spin index $ \alpha $ or $ \beta $. }
\label{SYKself}
\end{figure}

{\bf 5. $1/M $ corrections to two and four point correlation functions.  }
  Because of the conformably invariant form of Eq.\ref{detlaS}, we expect
  the propagator $  D ( \tau_1, \tau_2;  \tau_3, \tau_4 )  =  \langle \Sigma( \tau_1, \tau_2 ) \delta \Sigma( \tau_3, \tau_4 ) \rangle $
  takes also conformably invariant form \cite{naive}
  $ D ( \tau_1, \tau_2;  \tau_3, \tau_4 ) \sim 1/ | (\tau_1-\tau_3) ( \tau_2 - \tau_4 ) ( \tau_3-\tau_4 ) | $.
  Its contribution to self-energy at the order of $ 1/M $ was shown in Fig.2a:
\begin{eqnarray}
   \Sigma_{1/M}( \tau_1-\tau_4 ) & = & \int d \tau_2 d \tau_3  D ( \tau_1, \tau_2;  \tau_3, \tau_4 ) G_0( \tau_2 - \tau_3 )
                           \nonumber  \\
    & \sim &  1/ | \tau_1-\tau_4 |^{3/2}
\label{S1M}
\end{eqnarray}
  which takes the same scaling form as the saddle point $ \Sigma_{0}( \tau_1-\tau_4 ) $ at $ M= \infty $ in Eq.\ref{selfeq}.
  This indicates that the conformal invariance is kept at least to order of $1/M $.
  For example, it may change the coefficient $ \Lambda $ in Eq.\ref{break}, but not the function form such as
  the decay exponent $ 2 \Delta=1/2 $.
  We expect the conformal invariance is kept to all orders in $1/M $.
   In a sharp contrast, in the $ O(M) $ quantum rotor model, the $ 1/M $ corrections
   $  \Sigma( i \omega_n) \sim | \omega_n |^5 $ is more  subleading to the $ M= \infty $ result
   $ q^{aa}( i \omega_n) \sim | \omega_n | $.

  Its contribution to the $ Q^{ab} ( \tau- \tau^{\prime} ) $ function in Eq.\ref{saddle12}
  ( which is equal to the spin-spin correlation function )  at the order of $ 1/M $ was shown in Fig.2b:
\begin{eqnarray}
   Q_{1/M}( \tau-\tau^{\prime} )
    =  \int d \tau_1 d \tau_2 d \tau_3 d \tau_4 D ( \tau_1, \tau_2;  \tau_3, \tau_4 )     \nonumber  \\
    \times G_0( \tau - \tau_1 )G_0( \tau - \tau_3 ) G_0( \tau_2 - \tau^{\prime} ) G_0( \tau_4 - \tau^{\prime} )    \nonumber  \\
     \sim  1/ | \tau-\tau^{\prime} | ~~~~~~~~~~~~~~~~~~~~~~~~~~~~
\label{Q1M}
\end{eqnarray}
   which takes the same scaling form as that at $ M= \infty $ in Eq.\ref{saddle12}.
   It confirms the conformal invariance at least to order of $1/M $.

  In contrast to the SYK model which shows quantum chaos at the order $1/N $, here we fix at the $ N =\infty $ limit
  and perform a $ 1/M $ expansion, so in evaluating the OTOC Eq.\ref{otoc}, we need to take the same site index $ i=j $, but different
  $ O(M) $ component $ \alpha \neq \beta $ ( no sum over $ \alpha, \beta $ ):
\begin{equation}
  \langle \chi_{i \alpha} ( \tau_1) \chi_{i \alpha} ( \tau_2 ) \chi_{j \beta} ( \tau_3 ) \chi_{j \beta} ( \tau_4)  \rangle
\label{otoc}
\end{equation}
  which is essentially the extension of the $ Q^{ab} ( \tau, \tau^{\prime} ) $ function to 4 different times.

  When it is analytically continued to  $ \tau_1= 4 \beta/3, \tau_2=\beta/4, \tau_3= \beta/2 +it, \tau_4=it $ to compute
  the OTOC in real time  $ F_{\alpha \beta}(t)=  \langle Tr[ y \chi_{j \beta}(t) y \chi_{i \alpha} (0) y \chi_{j \beta}(t) y \chi_{i \alpha} (0)] \rangle $  with $ y^4=e^{-\beta H}/Z $. One may extract the Lyapunov exponent $ \lambda_L $.
  Unfortunately, just like $ Q^{ab} ( \tau- \tau^{\prime} ) $ in Eq.\ref{Q1M},
  it is still conformably invariant in the long time limit. Just like SYK, it may not show any quantum chaos at $ N =\infty $ and
  any $ M $. In order to study possible quantum chaos and evaluate the Lyapunov exponent $ \lambda_L $, one may need to study the $ 1/ N $ effects, then followed by
  the  $ 1/M $  expansion which be discussed below.

{\bf 6. $1/M $ corrections to the Free energy, zero temperature entropy and specific heat.  }
  From Eq.\ref{zn},\ref{z00P},\ref{z00}, we can evaluate the free energy per site and per spin component
  $ \beta f= \frac{F}{NM} $ at both $ N= \infty $ and  $ M=\infty $
  can be solely expressed in terms of the Green function:
\begin{equation}
  f_0= \frac{3J^2}{2} \int^{\beta}_{0} d \tau G^{4}( \tau ) -\frac{1}{2\beta} \sum_{i \omega_n} \log [- \beta G( i \omega_n) ]
\label{free00}
\end{equation}
  which can be used to evaluate the zero temperature entropy and specific heat,

 If plugging the conformally invariant solution Eq.\ref{break} into Eq.\ref{free00},
 one can see the first term just vanishes after regularizing the ultra-violet divergent integral properly,
 the second term leads to the zero temperature entropy $ s_0= -\partial f/\partial T $
 which was evaluated in the SY and the SYK model \cite{SY3,Kittalk,Mald}:
\begin{eqnarray}
  s_0 &= & \frac{1}{2} \log 2 - \pi \int^{\Delta}_0 (1/2-x) \tan \pi x   \nonumber   \\
    & = & \log2/8+ G/2 \pi =0.232424..
\label{s0}
\end{eqnarray}
  where $ G =0.916.. $ is the Catalan's constant.

  In order to evaluate the coefficient of the linear specific heat $ \gamma= C_v/T $ at low temperature, one need to consider the
  $ 1/\beta J $ correction to the conformally invariant solution Eq.\ref{break}. Then the first term in Eq.\ref{free00} does not vanish anymore,
  so it will also contribute under such a correction.

  In principle, using Eqs.\ref{detlaS}, one can evaluate the $ 1/M $ corrections to the free energy Eq.\ref{free00}.
  However, as alerted earlier, in contrast to evaluate the $ 1/M $ correction to the $ N $ point correlation functions,
  to get all the possible $1/M $ corrections to the free energy, one need also include the $ 1/M $ correction to
  $ Q(\tau,\tau^{\prime} ) $ in Eq.\ref{Q1M}. Because all these $1/M $ corrections are exactly marginal, so they
  will change the zero temperature entropy $ s_0 $ at $ M=\infty $ to become $ M $ dependent.
  It will not change the linear specific heat behaviour, but will make $ \gamma= C_v/T $ also $ M $ dependent.

{\bf 7. Instability to the QSG phase at $ N=\infty $ and a finite $ M $: }
 So far, we only focused  on the QSL phase, also ignored the replica off-diagonal fluctuations
in the QSL phase. It would be interesting to study if there is an instability to QSG order.
For $SU(M) $ SY model, it was argued in \cite{SY3} that the QSG still emerges as  the true ground state at any finite $ M $ below an
    exponentially suppressed temperature\cite{SY3} $ T_{QSG} \sim J e^{-\sqrt{M} } $.
    This is a non-perturbative effects which maybe inaccessible  to any orders in the $1/M $ expansion.

 To look at the QSG instability, as said below Eq.\ref{saddle1},  in the QSG phase, the replica off-diagonal  $ a \neq b $ component $  Q^{ab}( \tau- \tau^{\prime}  )= q \neq 0 $ which is independent of  $ \tau- \tau^{\prime} $.  If the replica symmetry is not broken, then $ q_{EA}= q $.
 So we split Eq.\ref{zn}  into the replica off-diagonal part and diagonal part:
 \begin{eqnarray}
 {\cal F}(Q) & = &  \frac{2(M-1)}{J^2 } \int d \tau  d \tau^{\prime} [ Q^{a \neq b} ( \tau, \tau^{\prime} )]^2 + \cdots  \nonumber   \\
 & - &  \log Z_0
 \label{znqsg}
\end{eqnarray}
  where the $ \cdots $ means the replica diagonal part. Then the $ Z_0 $ in  Eq.\ref{z0}  need to be replaced by:
\begin{eqnarray}
    Z_{0} & = &  \int {\cal D} \chi exp[ \frac{2}{M} \int d \tau  d \tau^{\prime} Q^{a \neq b} ( \tau, \tau^{\prime} )
       ( \chi^a_{\alpha} ( \tau) \chi^b_{\alpha} ( \tau^{\prime}) )^2    \nonumber  \\
       & + & \cdots ]
\label{z0qsg}
\end{eqnarray}
  After performing the cumulant expansion in Eq.\ref{z0qsg},  we can collect all the replica off-diagonal part into:
\begin{eqnarray}
    {\cal F}(Q^{a \neq b})  & = & \frac{2(M-1)}{J^2 } \int d \tau  d \tau^{\prime} [ Q^{a \neq b} ( \tau, \tau^{\prime} )]^2
                      \nonumber    \\
      & - & \frac{1}{2} [ \langle X^2 \rangle_{Z_{00}} - \langle X \rangle^2_{Z_{00}}]
\end{eqnarray}
  where $ X= \frac{2}{M} \int d \tau  d \tau^{\prime} Q^{a \neq b} ( \tau, \tau^{\prime} )
       ( \chi^a_{\alpha} ( \tau) \chi^b_{\alpha} ( \tau^{\prime}) )^2 $ is taking the average over the replica diagonal part
       of the single site/single component partition function $ Z_{00} $ in Eq.\ref{z00}.
       Finally, we reach:
\begin{widetext}
\begin{equation}
    {\cal F}(Q^{a \neq b})  =  \frac{2(M-1)}{J^2 } \int d \tau  d \tau^{\prime} [ Q^{a \neq b} ( \tau, \tau^{\prime} )]^2
      -4  \int d\tau_1 d\tau_2  d\tau_3 d\tau_4  Q^{a \neq b}( \tau_1, \tau_2 ) Q^{a \neq b}( \tau_3, \tau_4 )
     G^2_{0}( \tau_1 - \tau_3 ) G^2_{0}( \tau_2 - \tau_4 )
\label{qab}
\end{equation}
\end{widetext}
  whose structure  may be contrasted to Eq.\ref{detlaS}.

  Substituting the conformably invariant solution Eq.\ref{break} into Eq.\ref{qab} and regularizing the integral properly
  by introducing the natural dimensionless  short-time ( or high-energy ) cut-off $ \epsilon = 1/\beta J \ll 1 $, one reach
\begin{equation}
    {\cal F}(Q^{a \neq b})  = \frac{ 2 q^2 \beta^2 }{J^2} [ ( M-1 )- \frac{2}{\pi} \log^2(\beta J ) ]
\label{logbetaJ}
\end{equation}
  which leads to the QSG instability temperature
\begin{equation}
  T_{QSG}= J e^{- \sqrt{ \pi M/2 } }
\label{exp}
\end{equation}
 which shows this QSG instability is non-perturbative and in-accessible to any orders in the $ 1/M $ expansion.

 As shown above, the $ 1/M $ expansion preserves the conformally invariant form Eq.\ref{break}, so it will not change
 the exponential form Eq.\ref{exp} except it may modify the coefficient $ \pi/2 $ in Eq.\ref{exp}.
 Of course, due to the fermions can not condense,
 So the fermion Green function only has the replica diagonal saddle point solution, there is no QSG at $ M= \infty $.
 In fact, the QSG instability is non-perturbative, in-accessible to $ 1/M $ expansion to any orders.

{\bf 8. $ 1/N $ expansion at $ M=\infty $,  the Schwarzian action at both finite $ N $ and $ M $. }

In this work, we showed that at a fixed $ N=\infty $, the $ 1/M $ correction still keeps the long time conformal
or reparametrization invariance under $ \tau \rightarrow f(\tau) $ in Eq.\ref{ftau}.
So the result still applies to  $ O(3) $ case. However, it is not known if it applies to
the $ O(3) $ Heisenberg model due to the extra Majorana fermions and the associated $ Z_2 $ gauge field.
Then what would be the crucial quantum fluctuations effects from the $1/N $ effects ?
Hereㄛ we will derive the effective action at a finite $ N $, but with $ M=\infty $.
We find that it did not show any quantum chaos in this limit. We did not expect it to be. Because we
only expect quantum chaos show up at $1/N $ at any finite $ M $, which maybe explored in the $ 1/N $ expansion
followed by a $ 1/M $ expansion.

We expect the quantum chaos happen only in the spin singlet channel, so we ignore the quantum fluctuations
in spin symmetric and anti-symmetric channels \cite{spin}.
We also ignore the quantum fluctuations in the replica-off diagonal channel.
So we still assume that in both quantum spin liquid and quantum spin glass phase
 $  Q^{ab}_{\alpha \beta, \gamma \delta } ( \tau, \tau^{\prime} )
   = Q^{ab} ( \tau, \tau^{\prime} ) [ \delta_{\alpha  \delta } \delta_{\beta \gamma} - \delta_{\alpha \gamma  } \delta_{\beta \delta} ] $
   which is obviously anti-symmetric in $ (\alpha \beta) $ and $ (\gamma \delta ) $.
   Then Eq.\ref{zn} is simplified to:
\begin{eqnarray}
   \bar{ Z^n } = \int {\cal D} Q exp[ - {\cal F}(Q) ] ~~~~~~~~~~~~   \nonumber   \\
 \frac{ {\cal F}(Q)}{N} = \frac{2 (M-1) }{J^2 } \int d \tau  d \tau^{\prime} [ Q^{ab} ( \tau, \tau^{\prime} )]^2
 - \log Z_0
\label{znN}
\end{eqnarray}
 where the single site partition function $ Z_0 $ is:
\begin{eqnarray}
   Z_0 & = & \int {\cal D} P exp[ - M ( 2 \int d \tau  d \tau^{\prime}  Q^{ab}( \tau, \tau^{\prime})
   P^2_{ab}( \tau, \tau^{\prime} )      \nonumber    \\
       &- & \log Z_{00}   )  ]
\label{z00PN}
\end{eqnarray}
  where $ Z_{00} $ is the single-site and single-component partition function:
\begin{eqnarray}
    Z_{00} & = &  \int {\cal D} \chi exp[- \frac{1}{2} \int d \tau \chi^a_{\alpha} \partial_\tau \chi^a_{\alpha}
      + 4 \int d \tau  d \tau^{\prime} Q^{ab} ( \tau, \tau^{\prime} )    \nonumber  \\
      & \times & P_{ab}( \tau, \tau^{\prime} )
      \chi^a_{\alpha} ( \tau) \chi^b_{\alpha} ( \tau^{\prime}) ]     \nonumber   \\
      &=& Pf[ \partial_{\tau} \delta( \tau-\tau^{\prime} ) \delta^{ab}-\Sigma_{ab}( \tau, \tau^{\prime} ) ]
\label{z00N}
\end{eqnarray}
 where $ \Sigma_{ab}( \tau, \tau^{\prime} )= 8 Q^{ab} ( \tau, \tau^{\prime} ) P_{ab}( \tau, \tau^{\prime} ) $ is the self-energy.

    In the $ M \rightarrow \infty $ limit, $ P^{ab}( \tau, \tau^{\prime} ) $ takes the saddle-point value:
\begin{equation}
 P_{0ab}( \tau, \tau^{\prime} ) = G_{0ab}( \tau- \tau^{\prime} )= \frac{1}{M}
 \langle \chi^a_{\alpha} ( \tau) \chi^b_{\alpha} ( \tau^{\prime}) \rangle
\label{saddle2MN}
\end{equation}
   Then  Eq.\ref{z00PN} is simplified to
\begin{eqnarray}
  -\log Z_0 & = &  M [ 2 \int d \tau  d \tau^{\prime}  Q^{ab}( \tau, \tau^{\prime})
   P^2_{ab}( \tau, \tau^{\prime} )        \nonumber   \\
            & - & \log Z_{00}   ]
\label{z00PNs}
\end{eqnarray}
   and Eq.\ref{znN} is simplified to
\begin{eqnarray}
 \frac{ {\cal F}(Q)}{N M} & = &  \frac{2 }{J^2 } \int d \tau  d \tau^{\prime} [ Q^{ab} ( \tau, \tau^{\prime} )]^2
                            \nonumber   \\
    & + &  2 \int d \tau  d \tau^{\prime}  Q^{ab}( \tau, \tau^{\prime})
   P^2_{ab}( \tau, \tau^{\prime} )    \nonumber   \\
     & - &  \log Z_{00}
\label{1N}
\end{eqnarray}

  In Eq.\ref{1N}, the $ N $ and $ M $ appears in the combination $ N M $, so $ M \rightarrow \infty $
  also implies $ N M  \rightarrow \infty $ limit. As alerted earlier above Eq.\ref{znN},
  this is due to the fact that we have dropped the $ O(M) $ quantum spin fluctuations \cite{spin}.
  The physical limit should be $ N \rightarrow \infty $ first, then followed by $ M  \rightarrow \infty $
  to keep $ N \gg M $ instead of the other way around, so the order of limit may not commute.
  So we expect the quantum chaos may only happen in this physical limit instead of
  in the un-physical one. Both $ N $ and $ M $ need to be finite and  $ N \gg M $ to detect the possible quantum chaos.

  To be instructive, one may still take the saddle point of Eq.\ref{1N}
  $ \frac{\partial {\cal F}(Q)}{\partial Q }=0 $ which recovers Eq.\ref{saddle12}. Now we may
  substitute \cite{spin}
\begin{equation}
   Q^{ab}(\tau, \tau^{\prime} )= Q^{ab}_0(\tau- \tau^{\prime} ) + \delta  Q^{ab}(\tau, \tau^{\prime} )
\label{deltaQ}
\end{equation}
   into Eq.\ref{1N} and expand it to the quadratic order in $ \delta  Q^{ab}(\tau, \tau^{\prime} ) $.
   The zero-th order just leads to Eq.\ref{free00}. The first order vanishes due to the saddle point Eq.\ref{saddle12}.
   The quadratic order becomes:
\begin{eqnarray}
  \frac{ {\cal F}(Q)}{N M} & = &  \frac{2}{ J^2 } [ \int d\tau_1 d\tau_2  ( \delta Q( \tau_1, \tau_2 ) )^2
    +   \int d\tau_1 d\tau_2  d\tau_3 d\tau_4    \nonumber   \\
     & \times & \delta Q ( \tau_1, \tau_2 ) \delta Q ( \tau_3, \tau_4 )
     K_{1/N} ( \tau_1, \tau_2; \tau_3, \tau_4 ) ]
\label{zeroNM}
\end{eqnarray}
 where $ K_{1/N} ( \tau_1, \tau_2; \tau_3, \tau_4 )  = 8 J^2 G_{0}( \tau_1 - \tau_2 ) G_{0}( \tau_1 - \tau_3 )
   G_{0}( \tau_2 - \tau_4 ) G_0( \tau_3 - \tau_4 ) $. It is twice of the kernel in the $ 1/M $ expansion in Eq.\ref{detlas2}.
   By using the results achieved in the $ 1/M $ expansion below Eq.\ref{detlas2}, one can see Eq.\ref{zeroNM} is positive definite instead of having a zero mode \cite{three}. Just like the $ 1/M $ expansion at $ N=\infty $ presented in the previous sections, it only leads to conformably invariant  OTOC Eq.\ref{otoc} instead of an exponential growth \cite{spin}, so $ \lambda_L =0 $ at the leading order of the
   $ 1/M $ expansion.
   As argued above, it is not expected to appear at $ M=\infty $ anyway.

   Eq.\ref{zeroNM} can be contrasted to the quadratic order of the effective action $ S[ g, \sigma] $
   in the $ q=4 $ SYK Eq.4.3 in \cite{Mald}
   after integrating out $ g $:
\begin{eqnarray}
 \frac{ S[ \sigma ]_{SYK} }{N}  =   \frac{1 }{ 12 J^2 } \int d \tau  d \tau^{\prime} [  \sigma  ( \tau, \tau^{\prime} )]^2  ~~~~~~~~~~~~~~
                            \nonumber   \\
     +  \frac{1}{4} \int d\tau_1 d\tau_2  d\tau_3 d\tau_4
    \sigma ( \tau_1, \tau_2 ) \sigma ( \tau_3, \tau_4 )  K ( \tau_1, \tau_2; \tau_3, \tau_4 )
\label{zeroSYK}
\end{eqnarray}
   which contains a zero mode \cite{kernel} and otherwise positive definite.
   The lift of this zero mode by the irrelevant operator $ \partial_\tau $
   in Eq.\ref{selfeq} leads to the Schwarzian action in Eq.\ref{sch}.
   Note that here $ \delta \Sigma =3 J^2 G^2 \delta G \sim  G \sigma $,
   so $ \sigma \sim \delta G^2 $ which matches the $ \delta Q \sim \delta G^2 $ in Eq.\ref{zeroNM}.

    We expect that a zero mode and associated quantum chaos will show up
    in the physical limit $ N \rightarrow \infty $, then followed by a $ 1/M $ expansion.
    As shown in the previous sections, the $ 1/M $ expansion in Eq.\ref{deltaSigma}
       still keeps the  reparametrization invariance at any $ M $, so has very little effects except changing some
       coefficients ( such as the coefficient $ \Lambda $ in Eq.\ref{break} )
       in the conformally invariant solutions for 2- and 4- point correlation functions.
       Of course, the $ 1/M $ corrections will also change the kernel  $ K ( \tau_1, \tau_2; \tau_3, \tau_4 ) $,
       therefore its eigenvalues.
       Then the crucial expansion is $ 1/N $ which resembles the single $ 1/N $ expansion in the 4 indices SYK.
       When performing the $ 1/N $ expansion, one must expand around the true saddle point at $ N =\infty $
       valid at a finite $ M $ ( not Eq.\ref{deltaQ}  which holds only at $ N=\infty $ and $ M=\infty $ ), then using all the  2- and 4- point correlation functions, also the kernel
       $ K ( \tau_1, \tau_2; \tau_3, \tau_4 ) $ valid at this finite value of $ M $.
       They take the same functional forms as those in the 4 indices SYK, but the coefficients explicitly depend $ M $
       which, in principle, can be evaluated by the $ 1/M $ expansion outlined in the previous sections.


    At any $ N $ and $ M $,
    from Eq.\ref{zn},\ref{z00P},\ref{z00}, if dropping the kinetic  $ \partial_\tau $ term of the Majorana fermion
    in Eq.\ref{z00}, one can see that the action is invariant under the reparamatrization
    transformation $ \tau \rightarrow f(\sigma) $, the fermion field transforms as
     $ \chi^a_{\alpha} ( \tau ) \rightarrow [ f^{\prime}(\sigma)]^{-1/4} \chi^a_{\alpha} ( \sigma) $ and
    the HS fields transform as \cite{spin}:
 \begin{eqnarray}
     G(\tau_1,\tau_2) & = &[ f^{\prime}(\sigma_1) f^{\prime}(\sigma_2)  ]^{-1/4} G( \sigma_1,\sigma_2 ),   \nonumber   \\
    \Sigma(\tau_1,\tau_2) &= &[ f^{\prime}(\sigma_1) f^{\prime}(\sigma_2)  ]^{-3/4} \Sigma( \sigma_1,\sigma_2 ), \nonumber   \\
    Q(\tau_1,\tau_2) &= &  [ f^{\prime}(\sigma_1) f^{\prime}(\sigma_2)  ]^{-1/2} Q( \sigma_1,\sigma_2 )
\label{repara}
 \end{eqnarray}

    Because  the $ 1/M $ expansion still keeps the  reparametrization invariance\cite{symmetry} at any $ M $,
    the saddle point solution at any finite $ M $ spontaneously breaks the reparamatrization invariance
    to $ SL(2,R) $, leading to "zero mode " or Goldstone mode,
    while the irrelevant time derivative term explicitly breaks the re-parametrization symmetry and
    lifts the Goldstone mode to a pseudo-Goldstone mode whose quantum fluctuations
    can be described by  the Schwarzian in terms of $ f( \tau) $ re-parametrization.
\begin{equation}
 S[f]_{SYK/2}= - N \frac{\alpha_S(M)}{J} \int^{\beta}_0 d \tau \{ \tan \frac{ \pi f( \tau) }{\beta} , \tau \}
\label{sch}
\end{equation}
 where the Schwarzian is $ \{ f, \tau \} = \frac{ f^{\prime\prime\prime} }{ f^{\prime} }
       - \frac{3}{2} ( \frac{ f^{\prime\prime} }{ f^{\prime} } )^2  $.
        The coefficient is proportional to $ N $ and
        $ \alpha_S(M)= \alpha_S+ 1/M + \cdots $  which indicates that the
        quantum chaos shows up only at the order of 1 which is the sub-leading order in the $ 1/M $ expansion ( See the appendix ) \cite{subleading}.
        Namely, $ \lim_{M \rightarrow \infty} \alpha_S(M)= \alpha_S $ and  $ \lim_{M \rightarrow \infty} S[f]_{SYK/2}/M \sim N/M $.
        Note that $ f( \tau)=  \tau $ in Eq.\ref{sch} leads to a linear specific heat $ \gamma= C_v/T $ at low temperatures,
        which is the $ 1/M $ correction to that at $ M \rightarrow \infty $ presented below Eq.\ref{s0}.
        That is different from the SYK model where the Schwarzian action gives the same result \cite{Mald} as that from the free energy at $ N=\infty $.
        The difference is due to that only one $1/N $ expansion in the SYK,
        while here there is the $ 1/N $ expansion, followed by a $ 1/M $ expansion  \cite{subleading}.
        As stressed earlier, the two numbers $ N $ and $ M $ play different roles.
        Similarly, in the $ O(M) $ quantum rotor glass model, at a finite $ N $,  the quantum chaos
        does not happen in the leading order in the $ 1/M $ expansion, only appears  at the order of 1 which is in the sub-leading order
        in the $ 1/M $ corrections \cite{preliminary}.


The OTOC in Eq.\ref{otoc} is dictated by the Schwarzian and should show
similar behaviours as those of the OTOC in the 4 indies SYK:
there should be at least two time scales: $ t_d \sim \beta $ is the dissipation ( may also called relaxation time )
which is the characteristic time of Time ordered correlation function $ t_s \sim \beta \log N \gg t_d $ is the scrambling time
which is the characteristic time of OTOC. Both time may also depend on $ M $ which could also be a large number.
In the large $ N $ limit at any finite $ M $,
when $ t_d < t < t_s $, $ F(t)/F(0) =1 - \# e^{\lambda_L t }/N  $ where
$ \lambda_L $ is the Lyapunov exponent. At low temperatures $ 1 \ll \beta J \ll N $, $ \lambda_L = 2 \pi/\beta $ saturating the chaos upper bound.
At high temperatures $ \beta J \ll 1$, $ \lambda_L = J $.
When $ t \gg t_s $, it decays as a power law  $ F(t)/F(0) \sim t^{-6} $
as dictated by the 2d Liouville conformal field theory which reduces to the 1d Schwarzian when taking
the central charge $ c \rightarrow \infty $ limit
\cite{longtime1,longtime2,liu1,liu2}.
If so, the two indices SYK still show maximal chaos which may indeed fit the bulk string theory better than the four indices SYK.
Of course, at the order $1/N $, one may also need to consider
the $ O(M) $ spin fluctuations away from the saddle point Eq.\ref{saddle1}. We expect only spin singlet channel
shows the maximal chaotic behaviours, while the symmetric or ant-symmetric spin channel do not.

{\bf 9.  The QSG instability at a finite $ N $ }

When  $ \beta J  \gg N \gg 1 $, we expect the $ \log \beta J $ term in Eq.\ref{exp}
should be cutoff \cite{FS} by the finite size $ \log N $ , then the QSG instability in Eq.\ref{exp} happens
only when $ N > e^{\sqrt{\pi/2 (M-1) }} $. In the large $ N $ limit, followed by the large $ M $ limit,
if $ M < N < e^{\sqrt{\pi/2 (M-1) } } $, then the QSG can be safely avoided, the system remains in QSL
at $ T=0 $ and shows maximal chaos.
If $ M $ is sufficiently large, then there is a big such window.

 In fact, there could be also an intrinsic QSG instability in the SYK model.
 Indeed, recently, the SYK model was also argued to be  eventually a QSG phase at sufficiently low
 temperature $ T_{QSG} \sim J e^{-\sqrt{N} } $ ( note that here $ N $ is the number of sites in SYK, different than the $ M $ which is
 the $ SU(M) $ group in the SY )\cite{SYKsg1}, but later it was disputed in \cite{SYKsg2} by the following reason:
 in the conformally invariant limit $ N \gg \beta J \gg 1 $, there is indeed such an intrinsic QSG instability.
 However, in the strongly coupling limit $ \beta J \gg N \gg 1 $, the $ \log \beta J $ term should be cutoff by the $ \log N $, so
 the QSG instability disappears.
 Intuitively, in the strongly coupling limit $ \beta J \gg N \gg 1 $,
 the finite size effects are evident, the conformal invariance breaks down beyond the finite size of the system,
 one may not use the conformally invariant solution Eq.\ref{break} anymore.
 Any possible divergence must be cut-off by the finite size of the system.
 Similarly, the two and four point functions take
 different behaviours than the conformally invariant solution at a longer time beyond the finite size of the system \cite{longtime1,longtime2}.
 The biggest advantage of the 4 indices SYK is that there is only one large $ N $, so  the QSG instability automatically disappears.
 However, in the 2 indices SYK, there are two large numbers $ N $ and $ M $, which need to satisfy
 $ M < N < e^{\sqrt{\pi/2 (M-1) } } $ to avoid the QSG instability, then the QSL in this regime shows maximal chaos.

   Now we discuss the possible experimental realizations of the two indices SYK model.
   For the Majorana fermions,  the results achieved should still apply to the $ O(3) $ case.
   However, it is not known if it applies to
   the $ O(3) $ Heisenberg model due to the extra Majorana fermions and the associated $ Z_2 $ gauge field.
   Putting $ M=3 $ into $ M< N < e^{\sqrt{\pi/2 (M-1) } } $ leads to $ 3 < N < 5.88 $, which may be too small
   to show any quantum chaotic behaviours. While when $ N >5.88 $,  the system may fall into the QSG state instead of a QSL.
   For the complex fermions, the results achieved should still apply to the $ SU(2) $ case which is nothing but the
   random Heisengberg model Eq.\ref{SY}. Although it may be experimentally much more easily realized
   than its 4 indies counterpart \cite{coldwire}, it may fall into QSG instead of a QSL showing
   the maximal quantum chaos when $ N $ is sufficiently large such as $ N \sim 12-20 $.
   The energy level statistics, in both $ O(3) $ case and $ SU(2) $ case, in both bulk and especially the edge spectrum,
   will be studied in a separate publication \cite{yu} to see if they fall into QSG or show  $ N $ mod(8) ( or $ N_c  $ mod(4) ) Random Matrix
   pattern as the 4 indices  SYK did \cite{CSYKnum,MBLSPT,sff1}.

{\bf 10. Discussions and Conclusions. }


   As we showed here that the QSG may be always the ground state at $ T=0 $ ( namely below an exponentially suppressed temperature
   in the large $ M $ limit )  at $ N= \infty $. However, at a finite $ N $, when $ M < N < e^{\sqrt{\pi/2 (M-1) } } $,
   the finite size effect is dominating, so the QSG instability disappears.
   To some extent, this phenomenon maybe similar to the $ 1/N $ expansion in the $ U(1) $ Dicke model \cite{u1largen,gold,comment}
   which is also a $ 0+1 $ dimensional model.
   At sufficient large atom-photon coupling, at $ N=\infty $ limit, the normal phase
   will turn into the $ U(1) $ superradiant phase which breaks the global $ U(1) $ symmetry and leads to a zero ( Goldstone ) mode.
   However, at any finite $ N $, the super-radiant phase was washed away by the
   quantum phase diffusion process subject to a Berry phase in the imaginary time.
   The zero model was also lifted to a pseudo-Goldstone model with a finite energy scaling as $1/N $
   and a periodic dependence on the Berry phase.
   In fact, it maybe more similar to the $ 1/N $ expansion in the $ Z_2 $ Dicke model \cite{gold,comment,strongED}
   which is also a $ 0+1 $ dimensional model.
   The $ Z_2 $ superradiant phase breaks the global $ Z_2 $ symmetry and exists only
   at $ N=\infty $ limit. However, at any finite $ N $, the $ Z_2 $ super-radiant phase was washed away by the
   quantum tunneling process subject to a Berry phase between the two degenerate minima dictated by the $ Z_2 $ symmetry.
   Here at the thermodynamic limit $ N \rightarrow \infty $ limit, the system is frozen into the symmetry breaking
   QSG state when  $  T < T_{QSG} $ where it was trapped to one of infinite number of local minima landscape due to the quenched disorders,
   therefore breaks the ergodicity. The energy barriers between and two local minimum diverges  only at $ N=\infty $, but becomes finite at
   a finite $ N $.  So at any finite $ N $,  there are many instanton tunneling process among
   the infinite number of local minima  to recover the broken ergodicity, so the QSG state can be washed away by these instanton tunneling processes. If QSG can not be avoided, then it
   maybe interesting to study a possible replica symmetry breaking QSG at a finite $ N $.

 Another advantage over the 4 indices SYK is that it is easier to get to short-ranged models
( namely adding space dimensions )  just like ( but more complicated than ) the quantum rotor models in \cite{rotor1,rotor2,keldme}.
Now we confine $ J_{ij} $ to just nearest neighbour interaction in a $ d $ dimension cubic lattice.
Following the method in \cite{rotor2,keldme},
we can get a short-ranged model to include quantum fluctuations in $ d $ space dimensions \cite{preliminary}.
We will not only evaluate the Lyapunov exponent $ \lambda_L $, but also the butterfly velocity $ v_B $.

   In the original $ SU(M) $  fermionic SY model \cite{SY,SY3},
   due to the extra $ 1/M $ quantum fluctuations of the Lagrangian multiplier which is needed
   to fix the local boson or fermion constraint Eq.\ref{SYsum}, it is much more difficult to perform
   a direct $ 1/M $ expansion. However, by applying a local renormalization group analysis
   used in previous quantum impurity problems,
   the authors argued \cite{SY3,subir2} that the $ 1/M $ quantum fluctuations will not change the conformably
   invariant form of two and four point functions.
   They also argued in \cite{SY3} by looking at the QSG susceptibility that the QSG always emerges as
   the true ground state at any finite $ M $ below an
    exponentially suppressed temperature $ T_{QSG} \sim J e^{-\sqrt{M} } $.
   Here, in the context of the 2 indices Majorana SYK, we explicitly showed similar results by a direct $ 1/M $ expansion.
   Just like the 2 indices Majorana SYK studied here corresponds to the 4 indices Majorana SYK,
   the original SY model in its $ SU(M) $ representation \cite{SY}  can be called 2 indices complex fermion SYK,
   so it corresponds to the 4-indices complex fermion SYK \cite{CSYKnum} which has a global $ U(1) $ symmetry where
   a chemical potential term can be added to fix the total number of fermions \cite{tensorglobal}.
   We expect all the results achieved here should also apply to the 2 indices complex fermion SYK.
   Again, if QSG can be avoided, it may indeed fit the bulk string theory better than the four indices complex SYK.
   It would be interesting to look at its gravity dual also.  Ifthe  QSG can not be avoided, then it
   remains interesting to study how the replica symmetry breaking QSG explored at $ N=\infty $ in \cite{SY3}
   changes at a finite $ N $.

{\bf Acknowledgements }

 I thank Wenbo Fu for many helpful discussions and also his patient explanation of Ref.\cite{superSYK}.
 I also thank Song He for some general discussions and the careful reading of the manuscript.
 I am grateful for Yan Chen to invite me to deliver a lecture at the summer school 2018 at Fudan university.
 During this time at Fudan, I had fruitful discussions with
 S. Sachdev on several crucial questions addressed in this manuscript. I am also indebted to Sachdev for his careful reading of the
 manuscript and helpful comments. This research is supported by AFOSR FA9550-16-1-0412.

{\bf Appendix }

  In this appendix, we outline $ 1/N $ expansion followed by $1/M $ expansion.
  Putting $ N =\infty $ limit, we recover the $ 1/M $ expansion in Eq.\ref{detlaS}.
  Putting $ M =\infty $ limit, we recover the $ 1/N $ expansion in Eq.\ref{zeroNM}.
  However, at a finite $ N $ and a finite $ M $, we are not able to reach an analytical result, but
  we expect the sub-leading order in $ 1/M $, namely, at the order of $1/M^0 \sim 1 $, it should contain a zero mode
  which will be lifted by the irrelevant operator $ \partial_\tau/J $ to the Schwarzian Eq.\ref{sch}.

   At any finite $ N $, at a given $ Q^{ab} $ \cite{spin}, in the $ M \rightarrow \infty $ limit, $ P^{ab}( \tau, \tau^{\prime} ) $ takes the saddle-point value:
\begin{equation}
 P^0_{ab}( i \omega_n ) = \frac{1}{M}
 \langle \chi^a_{\alpha} ( \tau) \chi^b_{\alpha} ( \tau^{\prime}) \rangle_{Z^0_{00}}=( -i\omega_n  -\Sigma^0_{ab}( i \omega_n ) )^{-1}
\label{saddle2MNS}
\end{equation}
 where $ \Sigma^0_{ab}( \tau, \tau^{\prime} )= 8 Q^{ab} ( \tau, \tau^{\prime} ) P^0_{ab}( \tau, \tau^{\prime} ) $ is the self-energy
 at a given $ Q $.
 The single site/single component partition function $ Z^0_{00} $ is given in Eq.\ref{z00NS}.
 It  reduces to Eq.\ref{selfeq} only after putting $ N= \infty $.

 Note that Eq.\ref{saddle2MNS} indicates $ P^0_{ab}( \tau, \tau^{\prime} ) $ depends on $ Q^{ab} ( \tau, \tau^{\prime} ) $.
 If dropping the irrelevant $ \partial_{\tau} $ term in Eq.\ref{saddle2MNS}, in the conformal limit, one can write the dependence as:
\begin{equation}
   \int d \tau^{\prime}  P^0_{ab}(\tau- \tau^{\prime} ) 8 Q^{ab} ( \tau^{\prime}, \tau^{\prime \prime} ) P^0_{ab}( \tau^{\prime}, \tau^{\prime \prime} ) = -\delta(\tau- \tau^{\prime \prime } )
\label{deltaQS}
\end{equation}

   At the finite $ N $ and a finite $ M $, one can perform a $ 1/M $ expansion at a given $ Q^{ab} $ by writing:
\begin{equation}
   P^{ab}(\tau, \tau^{\prime} )= P^{ab}_0(\tau- \tau^{\prime} ) + \delta  P^{ab}(\tau, \tau^{\prime} )
\label{deltaP}
\end{equation}
   and expanding $ F(Q) $ in Eq.\ref{znN} to the quadratic  order in  $ \delta  P^{ab}(\tau, \tau^{\prime} ) $.
   Then  Eq.\ref{znN},\ref{z00PN},\ref{z00N}become:
\begin{eqnarray}
   \bar{ Z^n } & = & \int {\cal D} Q exp[ - {\cal F}(Q) ] ~~~~~~~~~~~~   \nonumber   \\
 \frac{ {\cal F}(Q)}{N} & = & \frac{2 (M-1) }{J^2 } \int d \tau  d \tau^{\prime} [ Q^{ab} ( \tau, \tau^{\prime} )]^2
 - \log Z_0            \nonumber   \\
 & - & \frac{1}{2} \int d \tau  d \tau^{\prime} \log Q^{ab} ( \tau, \tau^{\prime} )
\label{znNS}
\end{eqnarray}
 where the last term comes from the second HS transformation leading to  Eq.\ref{z00P}.
 This term was dropped in the previous literatures \cite{SY,SY3}, because they are not important
 in the $ N \rightarrow \infty $ limit, followed by the $ M \rightarrow \infty $ limit.
 However, as shown here, it may become important at a finite $ N $ and a finite $M $.

      The single site partition function $ Z_0 $ is:
\begin{eqnarray}
   Z_0 & = &  exp[ - M ( 2 \int d \tau  d \tau^{\prime}  Q^{ab} ( P^0_{ab} )^2-\log Z^0_{00} )]   \nonumber   \\
   & \times & \int {\cal D} \delta P exp[-M ( 2 \int d\tau_1 d\tau_2 Q^{ab}( \tau_1, \tau_2 ) ( \delta P_{ab}( \tau_1, \tau_2 ) )^2
    \nonumber   \\
   & + &  16  \int d\tau_1 d\tau_2  d\tau_3 d\tau_4
     Q^{ab} ( \tau_1, \tau_2 )  Q^{ab}( \tau_3, \tau_4 )    \nonumber   \\
    & \times &   P^0_{ab}( \tau_1- \tau_3 )  P^0_{ab}( \tau_2- \tau_4 )
      \delta P_{ab}( \tau_1, \tau_2 )\delta P_{ab}( \tau_3, \tau_4 ) ) ]
\label{z00PNS}
\end{eqnarray}
  where $ Z^0_{00} $ is the single-site and single-component partition function:
\begin{eqnarray}
    Z^0_{00} & = &  \int {\cal D} \chi exp[- \frac{1}{2} \int d \tau \chi^a_{\alpha} \partial_\tau \chi^a_{\alpha}
      + 4 \int d \tau  d \tau^{\prime} Q^{ab} ( \tau, \tau^{\prime} )    \nonumber  \\
      & \times & P^0_{ab}( \tau, \tau^{\prime} )
      \chi^a_{\alpha} ( \tau) \chi^b_{\alpha} ( \tau^{\prime}) ]     \nonumber   \\
      &=& Pf[ \partial_{\tau} \delta( \tau-\tau^{\prime} ) \delta^{ab}-\Sigma^0_{ab}( \tau, \tau^{\prime} ) ]
\label{z00NS}
\end{eqnarray}
 where $ \Sigma^0_{ab}( \tau, \tau^{\prime} )= 8 Q^{ab} ( \tau, \tau^{\prime} ) P^0_{ab}( \tau, \tau^{\prime} ) $ is the self-energy.
 listed below Eq.\ref{saddle2MNS}.

 In the $ M=\infty $ limit, setting  $ \delta P_{ab}=0 $ in Eq.\ref{z00PNS}, only the first line survives,
 then  Eq.\ref{znNS}  recovers the $ 1/N $ expansion at $  M= \infty $ in Eq.\ref{zeroNM}.

 In the $ N=\infty $ limit, using Eq.\ref{saddle12} and denoting $ \delta \sigma( \tau_1, \tau_2 )= J G_{0}( \tau_1, \tau_2 )  \delta P_{ab}( \tau_1, \tau_2 ) $ㄛ
 then Eq.\ref{z00PNS} recovers the $ 1/M $ expansion at $ N=\infty $ in Eq.\ref{detlas2}.

 Now at both finite $ N $ and finite $ M $, one must consider the $1/M $ quantum fluctuations $ \delta P_{ab} $.
 So integrating out $ \delta P_{ab} $ in Eq.\ref{z00PNS} leads to an effective potential for $ Q^{ab} $.
 Note that  $ P^0_{ab} $ also depends on $ Q^{ab} $ through Eq.\ref{saddle2MNS} or in the conformal limit
 through Eq.\ref{deltaQS}. If dropping the third and fourth line in Eq.\ref{z00PNS}, integrating $ \delta P_{ab} $ over  the second line
 leads to $ \frac{1}{2} \int d \tau  d \tau^{\prime} \log Q^{ab} ( \tau, \tau^{\prime} ) $ which just cancels
 the last term in Eq.\ref{znNS} due to the second HS transformation.
 This fact may hint that if taking into account the second, third and fourth lines, integrating $ \delta P_{ab} $ may
 lead to a zero mode at the order of 1 which is  at the subleading order in the $ 1/M $ expansion \cite{subleading}.
 Due to the dependence of  $ P^0_{ab} $  on $ Q^{ab} $ through Eq.\ref{saddle2MNS},
 It remains an open question to show it explicitly through a $ 1/N $ expansion at a finite $ M $.


\begin{thebibliography}{99}
\bibitem{om} Equivalently, one can choose the $ M(M-1)/2 $ generators of the $ O(M) $ group as
 $ L_{\alpha \beta}= -i \chi_{\alpha} \chi_{\beta}, \alpha < \beta $, it is easy to see that
 the total spin is $  \sum_{\alpha < \beta } L^2_{\alpha \beta}= M(M-1 )/8 $.   Setting $ M=3 $ also leads to
 spin $ s=1/2 $.
\bibitem{jacob} The Jacobian coming out of this transformation $ d \Sigma^{ab}/d P_{ab} = 8 Q^{ab}( \tau, \tau^{\prime} ) $
  will not affect the $ 1/M $ expansion. But as shown in the appendix,  any Jacobian could be important when considering
  both $ 1/N $ and $1/M $ expansion.
\bibitem{absolute} One can put the absolute value in $ G_{0}( \tau_1-\tau_2 ) $ and $ G_{0}( \tau_3-\tau_4 ) $
without affecting the results.
\bibitem{naive} Naively, if one just take the first term, then
the propagator $  D ( \tau_1, \tau_2;  \tau_3, \tau_4 )  =  \langle \Sigma( \tau_1, \tau_2 ) \delta \Sigma( \tau_3, \tau_4 ) \rangle $
 is simplified to be
  $ D ( \tau_1, \tau_2;  \tau_3, \tau_4 ) \sim \delta(\tau_1-\tau_3) \delta(\tau_2-\tau_4) / | (\tau_1-\tau_2) | $.
  The double line propagator collapses to a usual single line propagator.
  Then one can reach Eq.\ref{S1M},\ref{Q1M} immediately. Only the coeffiecient changes when one use the
  complete double line propagator form in Fig.1 and 2.
\bibitem{spin}  In principle, one need also to consider the quantum fluctuations in the spin anti-symmetric and symmetric channel
at any finite $ N $.
   In addition to the emergent local re-parametrization symmetry Eq.\ref{ftau},\ref{repara},
   there is also a emergent local $ O(M) $ symmetry, it is spontaneously broken down to the original global $ O(M) $ symemtry
   by the saddle point solution and  explicitly broken by the irrelevant $ \partial_\tau $ term. The resulting
   effective chiral field action \cite{polyakov} $ {\cal L}= Tr[ (g^{-1} \partial_{\tau} g )^2] $ where $ g \in O(M) $
   describes the quantum spin fluctuations \cite{highSYK1,liu2,tensorglobal}.
   So  the total effective low energy action contains  "Schwarzian" for the re-parametrization $ f( \tau ) $ in Eq.\ref{sch} plus
   the chiral field action. There could also be zero modes in the spin sector, but they will not lead to an exponential growth
   ( or quantum chaos ). If setting $ g= e^{i \phi} $ in the $ U(1) $ case,
   it leads to the quantum phase fluctuation  $ {\cal L}= ( \partial_{\tau} \phi  )^2 $ derived in \cite{subir3,highcSYK}.
   However, the $ U(1) $ sector does not lead to quantum chaos \cite{u1zero}.
\bibitem{symmetry}
    If the re-parametrization invariance at $ M =\infty $  were broken by the $ 1/M $ expansion, then as shown in \cite{Mald},
    there could be still quantum chaos, but not maximal anymore with $ \lambda_L < 2 \pi/\beta $.
\bibitem{kernel}  In \cite{Mald}, the kernel $ \tilde{K}( \tau_1, \tau_2; \tau_3, \tau_4 ) = -J^2 (q-1) K ( \tau_1, \tau_2; \tau_3, \tau_4 ) $
 has an eigenvalue $ 1 $. For $ q=4 $, it leads to the zero mode in Eq.\ref{zeroSYK}.
\bibitem{FS} In fact, one may view  $ \beta J $ as the finite size of the system along the imaginary time direction, while
$ N $ as the finite size of the system along the space direction. Then when $ \beta J < N  $ or $ \beta J > N  $,
$ \beta J $ or $ N $ can be used as the cutoff respectively. In principle, only the shortest length ( or the highest energy ) scale
can be used as a cutoff.
\bibitem{three} If it had been three times, it would have a zero mode at $ h=2, n=1 $.
\end{thebibliography}
\end{document}